\documentclass[twocolumn,aps,pra,showpacs,superscriptaddress,nofootinbib,10pt]{revtex4-1}
\usepackage[latin1]{inputenc}
\usepackage{graphicx,dcolumn,bm}
\usepackage{subfigure}
\usepackage{color}
\usepackage[colorlinks,hyperindex]{hyperref}
\usepackage{amsfonts}
\usepackage{amssymb}
\usepackage{amsmath}
\usepackage{latexsym}
\usepackage{times,txfonts}

\usepackage{mathrsfs}

\definecolor{Blue}{rgb}{0.0,0.0,1}
\definecolor{Red}{rgb}{1,0.0,0.0}
\definecolor{Green}{rgb}{0,0.5,0.0}

\begin{document}

\title{Thermal entanglement and teleportation in a dipolar interacting system}

\author{C.~S.~Castro}
\affiliation{Instituto de F\'{\i}sica, Universidade Federal Fluminense, Av. Gal. Milton Tavares de 
Souza s/n, Gragoat\'a, 24210-346, Niter\'oi, RJ, Brazil.}
\author{O.~S.~Duarte}
\author{D. P. Pires}
\author{D.~O.~Soares-Pinto}
\affiliation{Instituto de F\'{\i}sica de S\~ao Carlos, Universidade de S\~ao Paulo, P. O. Box 369, S\~ao Carlos, 
13560-970 SP Brazil.}
\author{M.~S.~Reis}
\affiliation{Instituto de F\'{\i}sica, Universidade Federal Fluminense, Av. Gal. Milton Tavares de 
Souza s/n, Gragoat\'a, 24210-346, Niter\'oi, RJ, Brazil.}

\date{\today }

\begin{abstract}
Quantum teleportation, which depends on entangled states, is a fascinating subject 
and an important branch of quantum information processing. The present work reports 
the use of a dipolar spin thermal system as a quantum channel to perform 
quantum teleportation. Non-locality, tested by violation of Bell's inequality and thermal entanglement, measured by negativity, unquestionably show that even entangled states that do not violate Bell's inequality can be useful for teleportation.
\end{abstract}


\maketitle

\section{Introduction}

Quantum mechanics is characterized by its counter-intuitive concepts as, for example, 
quantum entanglement, whose importance in modern physics has stimulated intensive research 
of several quantum systems~\cite{Horodecki:09,Amico:08}. Entanglement theory treats the 
quantum correlations among states of a system which may lead to observation of non-local phenomena 
like the violation of Bell's inequality~\cite{Brunner:14}. Although its great importance, there is not 
a general operational criterion necessary and sufficient to determine whether a state of arbitrary 
purity and dimension is entangled or not~\cite{Amico:08,Horodecki:09}. However, in a some bipartite 
systems, like those described in $2 \otimes 2$ and $2 \otimes 3$ Hilbert spaces, several measures, 
such as negativity, are available for quantifying entanglement~\cite{2007_QuantInfComput_1_51_PlenioVirmani}. In terms of quantum information 
purposes, quantum entanglement is an important resource. First, entanglement of a bipartite 
system represents a fundamental requirement to allow quantum teleportation, which is an information 
protocol for transmitting an unknown state from one place to another without a physical 
transmission channel. Such protocol can be seen as a corner stone of quantum information processing 
because it was the first one showing the usefulness of quantum 
entanglement~\cite{Bennett:93}. On the other hand, thermal entanglement, i.e.,
the entanglement of quantum systems at finite temperature, is one of the links between quantum 
information and condensed matter areas~\cite{Amico:08}, and consequently, it has been extensively 
studied by both, theoretical and experimental physicists~\cite{Souza:08,Souza:09,Duarte:13,
DuarteEPL:13,Rappoport:09,Physica:10,Candini:10,Troiani:13,Troiani:14}.

In this work we report the effects of a dipolar interaction between two spins on 
their degree of entanglement and nonlocality. Also, considering such model a quantum communication channel, we analyse the effects of such interaction over its communication capacity throught the teleportation fidelity. Such interaction arises 
due to the influence of the magnetic field created by one magnetic moment on the site of another 
magnetic moment~\cite{Reis:13}. We begin with the model of dipolar interaction and show that, for 
the case of two coupled spins $1/2$, whatever is the ground state, we have the presence of entanglement. We quantify the amount of 
entanglement by using negativity and verify that our model presents some degree of entanglement at a 
given coupling parameters $\Delta$ and $\epsilon$. In adition, we show how the magnetic anisotropies can influence the process 
of teleportation, which is based on the degree of entanglement of the quantum states involved in the 
process. We calculate the averaged teleportation fidelity and verify that this quantity has a similar 
behavior of negativity and violation of Bell's inequality. Such process successfully occurs without need of nonlocality of quantum states~\cite{Hardy:99,Bschorr:01,Barrett:01,Cavalcanti:13,Popescu:94}.

%%%%%%%%%%%%%%%%%%%%%%%%%
\section{The Model}
\label{model}
%%%%%%%%%%%%%%%%%%%%%%%
The dipolar interaction arises from the magnetic field created by a magnetic moment of a spin $\vec{\mu}=-\mu_{B} 
g \vec{S}$ ~\cite{Reis:13}, where $\mu_B$ is the Bohr's magneton and $g$ the giromagnetic factor, on the site of another spin and it is represented by the Hamiltonian
\begin{equation}
 \label{eq:hamiltonian001}
H = -\frac{1}{3}{\vec{S}_1^{T}}\cdot\tensor{T}\cdot{\vec{S}_2} ~,
\end{equation}
where $\tensor{T} = \text{diag}(\Delta - 3\epsilon,\Delta + 3\epsilon,-2\Delta)$ 
is a diagonal tensor, ${\vec{S}_j} = \{ {S_j^x},{S_j^y},{S_j^z}\}$ is the spin operator, $\Delta$ and 
$\epsilon$ are the dipolar coupling constants between the spins. In addition, $\Delta < 0$ 
indicates that spin is on $x-y$ plane, while $\Delta > 0$ indicates that spin is on $z$-axis.

That hamiltonian can describe a pair of spin $1/2$ particles and can be written in a matrix form
\begin{equation}
\label{Hmatrix}
H = \frac{1}{6} \left(
\begin{array}{cccc}
 \Delta & 0 & 0 & 3\epsilon \\
0 & - \Delta & - \Delta & 0 \\
0 & - \Delta & - \Delta & 0 \\
3\epsilon & 0 & 0 &  \Delta
\end{array}\right) ~,
\end{equation}
with the respective eigenvalues and eigenvector given by
\begin{equation}
 \label{eq:energies0001}
{\mathcal{E}_{\Psi^-}} = 0 ~,\quad {\mathcal{E}_{\Psi^+}} = -\Delta/3 ~,\quad
{\mathcal{E}_{\Phi^{\mp}}} = (\Delta \mp 3\epsilon)/6 ~.
\end{equation}
\begin{equation}
 \label{bellstates}
|\Psi^{\pm}\rangle = \frac{1}{\sqrt{2}}(|01\rangle \pm |10\rangle) ~,\quad 
|\Phi^{\pm}\rangle = \frac{1}{\sqrt{2}}(|00\rangle \pm |11\rangle) ~,
\end{equation}
Note that the eigenvectors are the four Bell states which are well known bipartite entangled pure states. This hamiltonian can be written through the spectral decomposition in terms of its states, i.e., $H = {\sum_{\alpha}}{\mathcal{E}_{\alpha}}|\alpha\rangle\langle\alpha|$, where $\alpha \in \{{\Phi^{+}},{\Phi^{-}},{\Psi^{+}}, {\Psi^{-}}\}$. 

Let's consider the system at thermal equilibrium described by the canonical ensemble $\rho = {\mathcal{Z}^{-1}}{e^{-H/{k_B}T}}$, where 
$H$ is the hamiltonian of the system, ${k_B}$ the Boltzmann's constant and $\mathcal{Z} = \text{Tr}(e^{-H/{k_B}T})$ 
the partition function. Since $\rho$ is a thermal operator, the entanglement on this state is 
called thermal entanglement~\cite{Wang:01,Arnesen:01,Nielsen:00}. From using Eq.~\eqref{Hmatrix}, we obtain the density operator,
\begin{equation}\label{eq:densitymatrix}
\rho = \left(
\begin{array}{cccc}
\rho_{11} & 0 & 0 & \rho_{14} \\
0 & \rho_{22} & \rho_{23} & 0 \\
0 & \rho_{23} & \rho_{22} & 0 \\
\rho_{14} & 0 & 0 & \rho_{11} \\
\end{array}\right) ~,
\end{equation}
where
\begin{eqnarray}\label{eq:elementos}
\rho_{11} &=& \frac{1}{\mathcal{Z}}{e^{-{\beta \Delta}/6}}\cosh\left(\frac{\beta\epsilon}{2}\right) ~, \nonumber \\
\rho_{22} &=& \frac{1}{\mathcal{Z}}{e^{{\beta \Delta}/6}}\cosh\left(\frac{\beta \Delta}{6}\right) ~, \nonumber \\
\rho_{23} &=& \frac{1}{\mathcal{Z}}{e^{{\beta \Delta}/6}}\sinh\left(\frac{\beta \Delta}{6}\right) ~,\nonumber \\
\rho_{14} &=& -\frac{1}{\mathcal{Z}}{e^{-{\beta \Delta}/6}}\sinh\left(\frac{\beta\epsilon}{2}\right)
\end{eqnarray}
and
\begin{equation}
\mathcal{Z} =2{e^{{\beta \Delta}/6}}\cosh\left(\frac{\beta \Delta}{6}\right) + 
2{e^{-{\beta \Delta}/6}} \cosh\left(\frac{\beta\epsilon}{2}\right)~.
\end{equation}
The parameters $\Delta$ and $\epsilon$ can be experimentally determined since 
they are within the partition function $\mathcal{Z}$ and, consequently, within the thermodynamical quantities like magnetic susceptibility and heat capacity.

The density operator can also be written in terms of Bell states as $\rho = {\sum_{\alpha}}
{p_{\alpha}}|\alpha\rangle\langle\alpha|$, where ${p_{\alpha}} = {\mathcal{Z}^{-1}}{e^{-{\mathcal{E}_{\alpha}}
/{k_B}T}}$ is the Boltzmann weight and $\alpha \in \{{\Phi^{+}},{\Phi^{-}},{\Psi^{+}}, {\Psi^{-}}\}$. Note that the density matrix of the system is expressed in terms of the four Bell states, which are not possible to be written as a convex sum of the original spin states. Thus, the system will always present 
some degree of entanglement for any of them as the ground state, except when two 
or more of these states present the same occupation probability, in that case the ground state will be a mixture that produce a separable state.

Generally speaking, the density operator  can be written in the Fano form~\cite{Horodecki:95}
\begin{equation}
\label{eq:rhot}
\rho = \frac{1}{4}\left[\mathbb{I}\otimes\mathbb{I} + \vec{r}\cdot\vec{\sigma}\otimes\mathbb{I} + 
\mathbb{I}\otimes\vec{s}\cdot\vec{\sigma} + {\sum_{i,j}}{c_{ij}}{\sigma_i}\otimes{\sigma_j}\right] ~,
\end{equation}
where $\mathbb{I}$ is the $2 \times 2$ identity operator, ${\sigma_i}$ are the Pauli 
matrices, ${r_j} = \langle{\sigma_j}\otimes\mathbb{I}\rangle$, ${s_j} = \langle\mathbb{I} 
\otimes{\sigma_j}\rangle$, and $c_{ij} = \langle {\sigma_i} \otimes {\sigma_j} \rangle$ 
are spin-spin correlation functions. For our case Eq.~\eqref{eq:densitymatrix}, 
${r_j} = 0$, ${s_j} = 0$, ${c_{ij}} = {\delta_{ij}}{c_i}$ and then
\begin{equation}
\label{eq:rhobelldiag}
\rho = \frac{1}{4}\left(\mathbb{I}\otimes\mathbb{I} + {\sum_i}{c_i}{\sigma_i}\otimes 
{\sigma_i}\right) ~,
\end{equation}
where
\begin{eqnarray}\label{eq:qsm004}
{c_1} &=& \frac{2}{\mathcal{Z}} \left[{e^{{\beta  \Delta}/6}}\sinh\left(\frac{\beta  \Delta}{6}\right) 
- {e^{-{\beta  \Delta}/6}}\sinh\left(\frac{\beta\epsilon}{2}\right)\right] ~, \nonumber \\
{c_2} &=& \frac{2}{\mathcal{Z}}\left[{e^{{\beta  \Delta}/6}}\sinh\left(\frac{\beta  \Delta}{6}\right) 
+ {e^{-{\beta  \Delta}/6}}\sinh\left(\frac{\beta\epsilon}{2}\right) \right] ~, \nonumber \\
{c_3} &=& \frac{2}{\mathcal{Z}} \left[-{e^{{\beta  \Delta}/6}}\cosh\left(\frac{\beta  \Delta}{6}\right) 
+ {e^{-{\beta  \Delta}/6}}\cosh\left(\frac{\beta\epsilon}{2}\right)\right] ~.
\end{eqnarray}
The above density operator is named Bell-diagonal state and the coefficients $c_{i}$ compose a diagonal correlation matrix $\mathcal{C}$. This general form of the density matrix is extremely useful to explore the role of correlations on the system as will be seen in the next section. 

%%%%%%%%%%%%%%%%%%%%%%%%%%%%%%%%%%%%%%%
\section{Entanglement and Nonlocality}
\label{entangnonlocal}
%%%%%%%%%%%%%%%%%%%%%%%%%%%%%%%%%%%%%%%

\subsection{Nonlocality}

Bell's inequality, which imposes an upper limit on the correlation between measurements 
made on observables of separable qubits, is used here to detect non-locality in our system 
of dipolar interaction. Such inequality states that, 
in the absence of non-local effects, the correlation between measurements made on two qubits 
should be limited by $\pm2$. However, quantum mechanics imposes limits of $\pm 2\sqrt{2}$ 
on the same quantities for pure entangled states~\cite{Brunner:14}. Consequently, violation of Bell's inequality 
is directly related to non-local entangled states~\cite{Acin:06,Brukner:04}. 

Probably, the most known Bell's inequality is that of Clauser, Horne, Shimony, and Holt 
(CHSH)~\cite{Clauser:69}, which can be tested experimentally~\cite{Fox:06}. The Bell operator associated with the CHSH quantum inequality is
\begin{equation}
B_{CHSH} = \vec{a} \cdot \vec{\sigma} \otimes (\vec{b} + \vec{b'}) \cdot \vec{\sigma} + \vec{a'} 
\cdot \vec{\sigma} \otimes (\vec{b} - \vec{b'}) \cdot \vec{\sigma} ~,
\end{equation}
where $\vec{a}$, $\vec{a'}$, $\vec{b}$, and $\vec{b'}$ are unit vectors in $\mathbb{R}^{3}$, and 
the CHSH inequality is therefore
\begin{equation}
|\langle B_{CHSH} \rangle| \le 2 ~.
\label{inequality}
\end{equation} 
If this inequality is violated, the system is non-locally entangled. However, $\langle B_{CHSH}\rangle$ 
depends on the chosen directions. Therefore, the CHSH inequality can be intentionally violated by 
choosing the directions that maximize $\langle B_{CHSH}\rangle$. This procedure is defined as
\begin{equation}\label{eq:bell}
\mathcal{B} = \mbox{max} |\langle B_{CHSH} \rangle |.
\end{equation}
To determine $\mathcal{B}$, Horodecki and coworkers~\cite{Horodecki:95} proposed a necessary and sufficient condition for a mixed state of two spins $1/2$  in the Bell-diagonal form to violate the CHSH inequality. One first define the matrix $U = {\mathcal{C}^{T}}\mathcal{C}$, where 
$\mathcal{C}$ is the correlation matrix and $\mathcal{C}^{T}$ is its transpose.  The quantity  $M(\rho) = {\max_{i<j}}({u_i} + {u_j})$  is defined~\cite{Horodecki:95}, where $u_{i}$ are the eigenvalues of the matrix $U$ and the 
Eq.~\eqref{eq:bell} can be written as $\mathcal{B}(\rho) = 2\sqrt{M(\rho)}$.

It is straightforward to verify that for the present case the matrix is $U = \text{diag}({c_1^2},
{c_2^2},{c_3^2})$. From the eigenvalues of $U$ we obtained $M$ for our system of 
dipolar interaction and thus the analytical expression for $\mathcal{B}$ is
\begin{equation}
\mathcal{B}(\rho) =2\sqrt{\max\{{c_1^2} + {c_2^2}, {c_1^2} + {c_3^2}, {c_2^2} + {c_3^2}\}} ~.
\end{equation}
It is worth remember that any violation of the inequality $\mathcal{B}(\rho)\le 2$ implies non-locality. In Fig.~\ref{Bell} we can see $\mathcal{B}(\rho)$ as function of $\Delta/{k_B}T$ and $\epsilon/{k_B}T$. The red line is the boundary given by $\mathcal{B}(\rho) = 2$ and thus the states inside the contour defined by the red line are the ones that do not violate the Bell's inequality test, and therefore are local quantum states. It is easy to see that as higher is the ratio $\Delta/{k_B}T$ and $\epsilon/{k_B}T$,  more states violate Bell's inequality.

%%%%%%%%%%%%%%%%%
\subsection{Entanglement}
%%%%%%%%%%%%%%%%%

In general, partial transposition does not retain the positivity required from 
any density operator, unless, as proven by Horodecki {\it et al}~\cite{Horodecki:96}, the 
state is separable. The concept of negativity~\cite{Werner:89,Vidal:02} is derived from the 
Peres-Horodecki separability criterion~\cite{Horodecki:96,peres:96} which states that for 
systems with Hilbert space dimension $2 \otimes 2$ and $2 \otimes 3$, there is entanglement 
if, and only if, its partial transpose is not positive definite. Thus, the negativity 
is defined as~\cite{Vidal:02,Audenaert:00}
\begin{equation}
\mathcal{N}(\rho) = \frac{{\|{\rho^{T_A}}\|_1} - 1}{2} ~,
\label{neg}
\end{equation}
where ${\|A\|_1} = \text{Tr}\sqrt{{A^{\dagger}}A}$ is the trace norm, and ${\rho^{T_A}}$ is the partial transposition 
of the density operator $\rho$ whose elements are $\langle\alpha\beta|{\rho^{T_{A}}}|\gamma\delta\rangle = 
\langle\gamma\beta|\rho|\alpha\delta\rangle$. As can be proved, negativity is convex and does not increase 
under local operations and classical comunication (LOCC), i.e., it is an entanglement monotone~\cite{Vidal:02}. 
Moreover, negativity is normalized such that $0 \le \mathcal{N} \le 1$, where $\mathcal{N} = 0$ denotes 
completely separable and $\mathcal{N} =1$ denotes maximally entangled states~\cite{Vidal:02}. For the present model, Eq.~\eqref{eq:densitymatrix}, the partial 
transposition gives
\begin{equation}
{\rho^{T_A}} = \left(\begin{matrix}
\rho_{11} & 0 & 0 & \rho_{23} \\
0 & \rho_{22} & \rho_{14} & 0 \\
0 & \rho_{14} & \rho_{22} & 0 \\
\rho_{23} & 0 & 0 & \rho_{11} \\
\end{matrix}\right) ~,
\end{equation}
and so the negativity reads as 
\begin{equation}
 \label{eq:negat001}
\mathcal{N}(\rho) = \frac{1}{4}\left(|c_{1} + c_{3}| + |1 + c_{2}| + 
|c_{1} - c_{3}| + |1 - c_{2}| - 1\right) ~.
\end{equation}

\begin{figure}
\centering
\subfigure[ref1][]{\includegraphics[scale=0.22795]{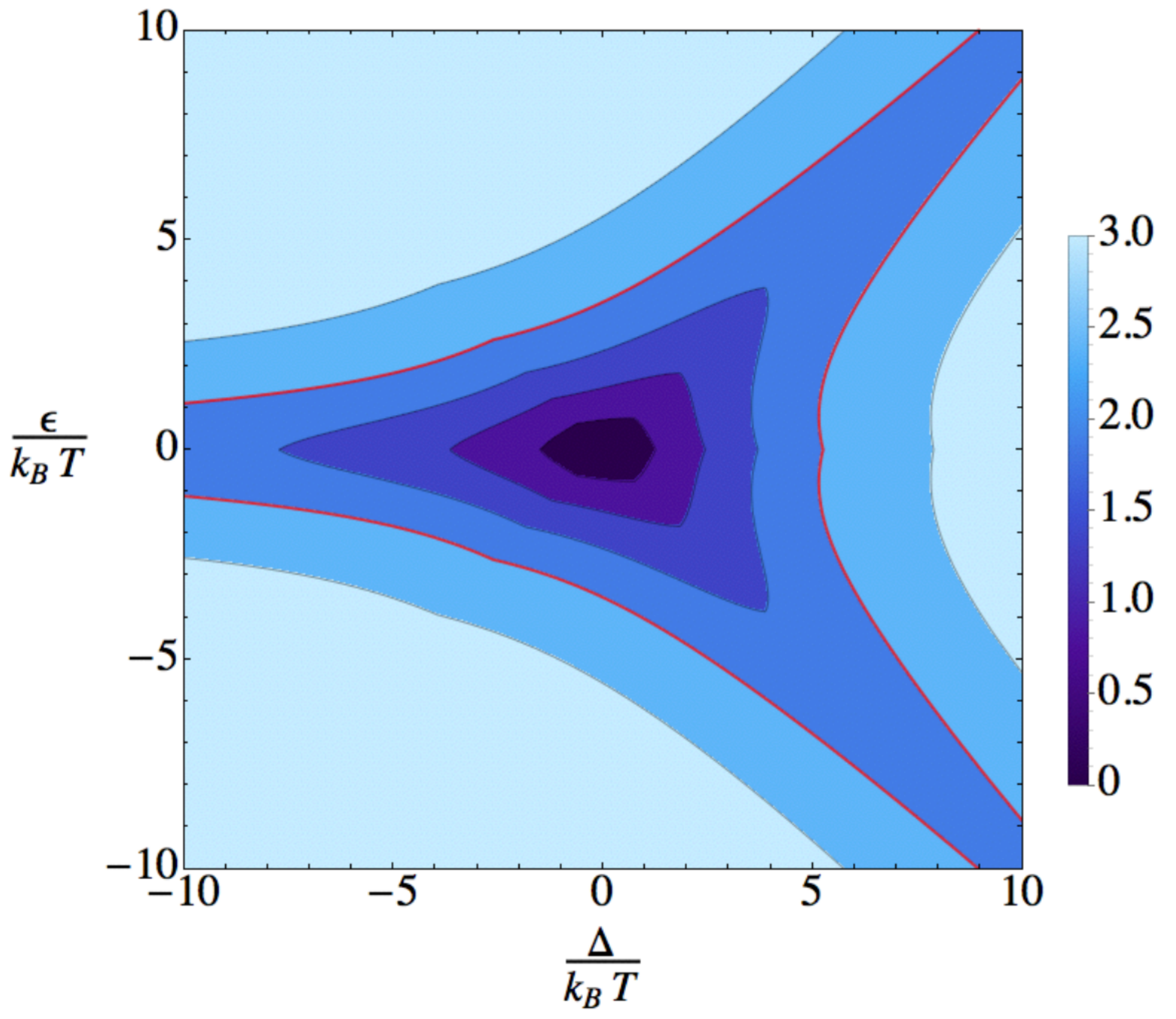}\label{Bell}}
\subfigure[ref2][]{\includegraphics[scale=0.22795]{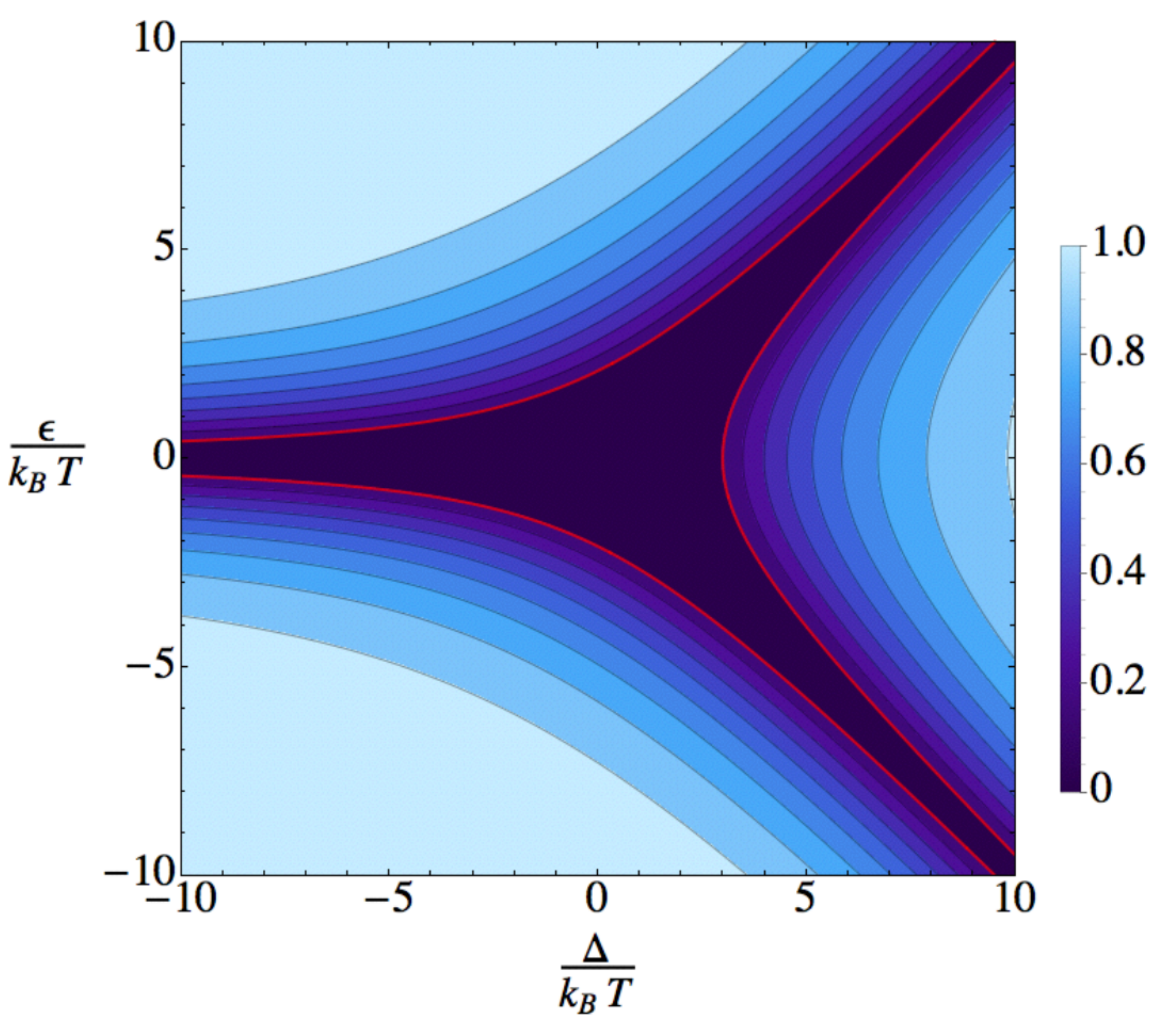}\label{Negativity}}
\subfigure[ref3][]{\includegraphics[scale=0.22795]{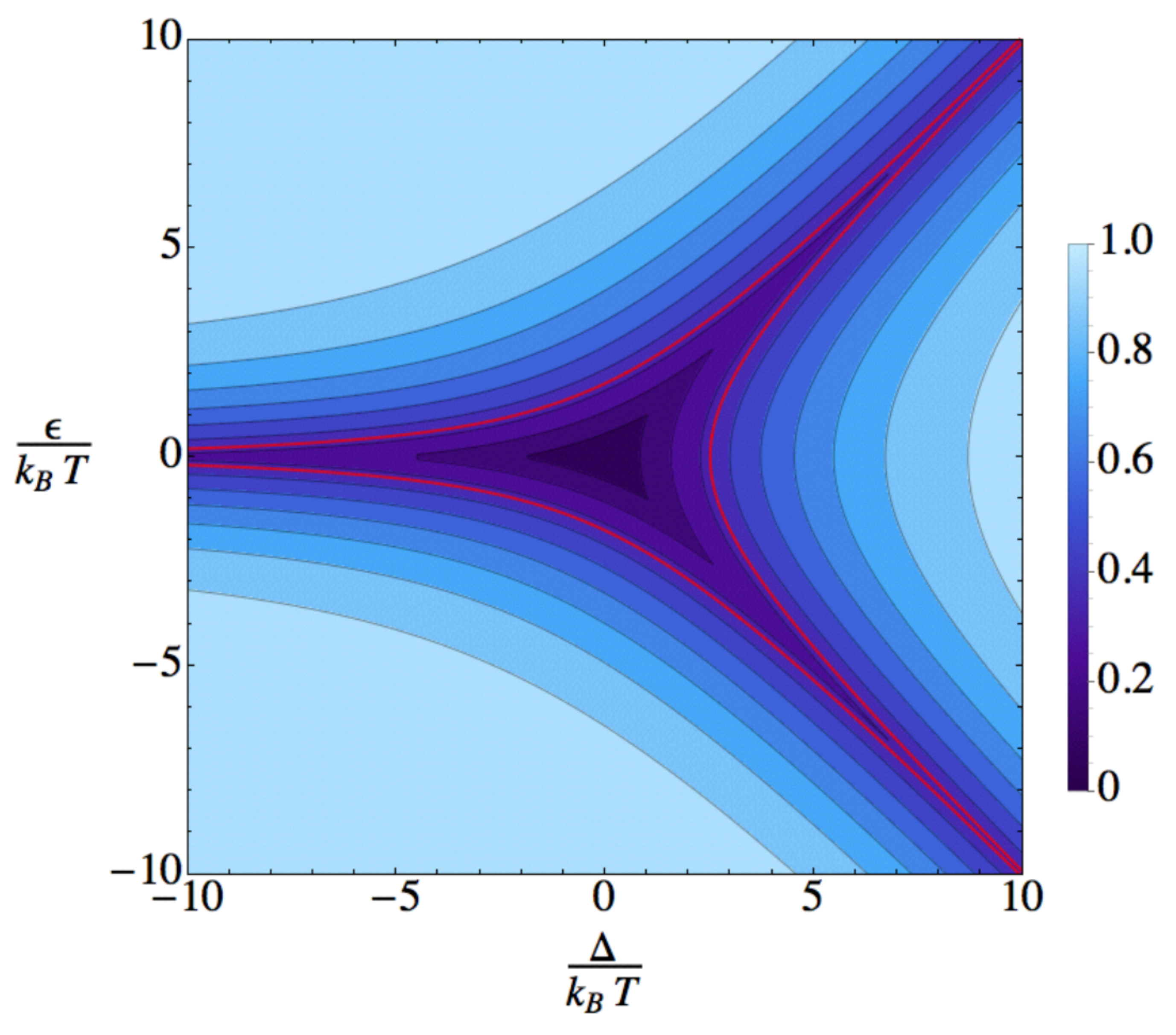}\label{Fidelity}}
\caption{(Color on line) Bell's inequality violation test $\mathcal{B}(\rho)$ (Fig.~\ref{Bell}) and Negativity $\mathcal{N}(\rho)$ (Fig.~\ref{Negativity})
as function of  $\Delta/{k_B}T$ and $\epsilon/{k_B}T$. The violation of Bell's inequality occurs in regions where 
$\mathcal{B} > 2$ and, thus, in this region there is a non local thermal state. Maximum entanglement occurs for values in which  $\epsilon \to 0$ and $\Delta \gg {k_B}T$ or $\Delta < 0$ and $|\epsilon| \gg {k_B}T$. Fidelity of teleportation $\mathcal{F}$ (Fig.~\ref{Fidelity}) as function of $\Delta/k_{B}T$ and $\epsilon/k_{B}T$. 
As expected region for $\mathcal{F}$ has the same behaviour of figures of negativity and Bell's inequality violation test. The red lines are the boundary given by $\mathcal{B}(\rho) = 2$, $\mathcal{N}(\rho) = 0$, and ${\mathcal{F}_{min}} = 2/3$, the last one being the minimal value to successfully occur the quantum teleportation process.}
\label{figall}
\end{figure}

Fig.~\ref{Negativity} shows $\mathcal{N}(\rho)$ as function of $\Delta/{k_B}T$ and $\epsilon/{k_B}T$. The red line is the boundary given by $\mathcal{N}(\rho) = 0$ and thus the states inside the contour defined by the red line are the non entangled ones. Again, as higher is the ratio $\Delta/{k_B}T$ and $\epsilon/{k_B}T$, more entanglement is present in the system. Comparing the figures for $\mathcal{N}(\rho)$ and $\mathcal{B}(\rho)$ we see that for our system, although both present the same behaviour, not all entangled states are non local in the sense of violating Bell's inequality. Thus, one question may be raised: Considering this system in thermal equilibrium a channel that can promote teleportation, what would be its source for quantum communication capability? Next, section we will discuss details about that question.
\begin{figure}[!h]
\centering
\includegraphics[scale=0.22795]{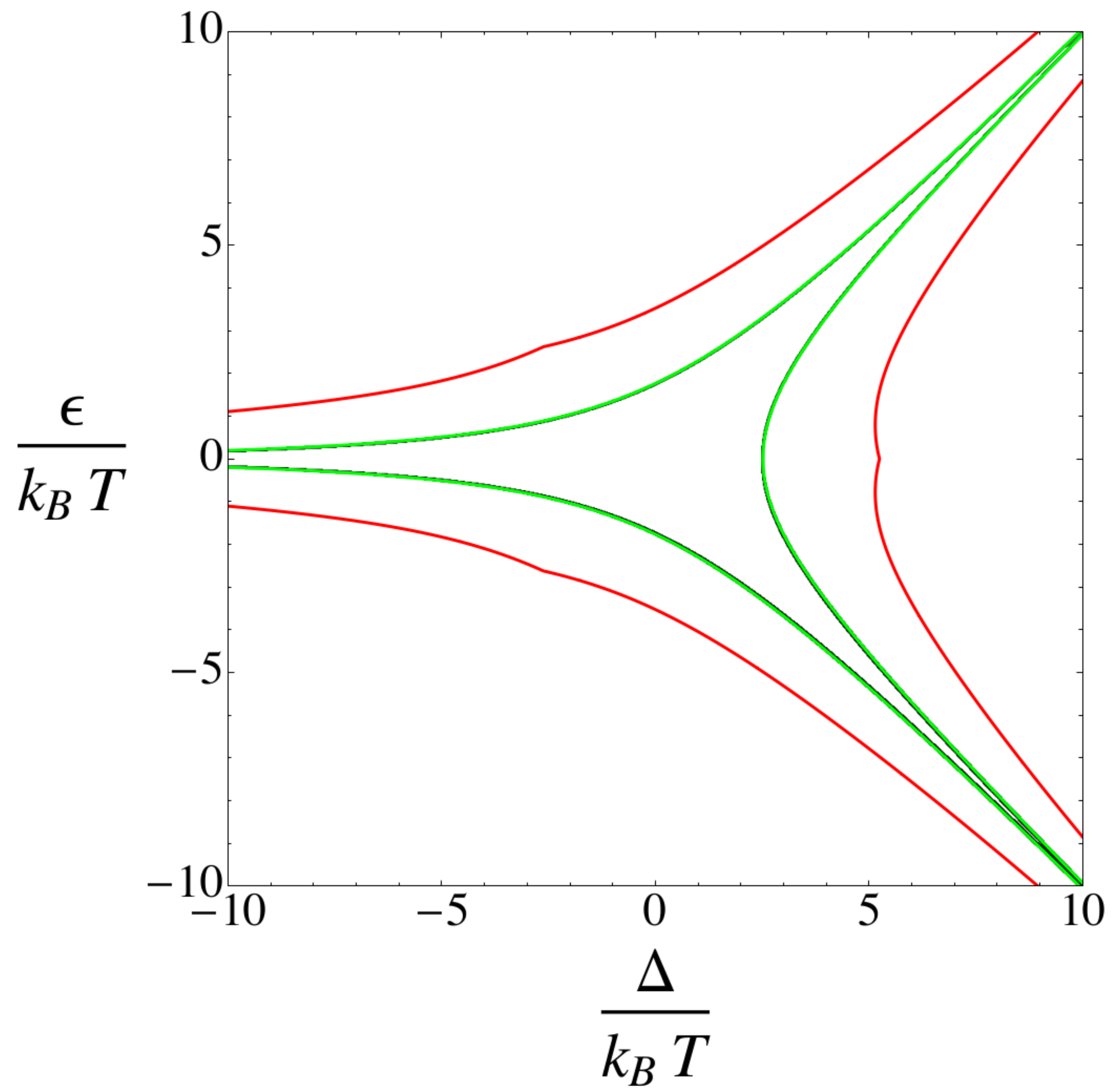}
\caption{(Color on line) Superposition of the red lines in figures ~\ref{figall} in terms of $\Delta/k_{B}T$ and $\epsilon/{k_B}T$. The red line is to refer Bell ($\mathcal{B}(\rho)=2$), the black line is to refer Negativity ($\mathcal{N}(\rho)=0$), and the green line is to refer Fidelity ($\mathcal{F}(\rho)=2/3$). In this figure is emphasized that all the entangled states are useful for teleportation.}
\label{superposition}
\end{figure}

%%%%%%%%%%%%%%%%%%%%%%%%%%%%%%%%%%%%%%
\section{Teleportation}
\label{fidelitysection}
%%%%%%%%%%%%%%%%%%%%%%%%%%%%%%%%%%%%%

The basic idea of quantum teleportation is to transfer a unknown quantum state of 
a particle to a separated one without a physical communication channel. The seminal 
proposal of Bennet and co-workers presents a teleportation protocol based on Bell 
measurements, which are projective measurements over one of the four states of the 
Bell basis (see Eq.~\eqref{bellstates}), and Pauli matrices 
rotations. The protocol can be described as follows~\cite{Bowen:01}. Consider a pair 
of particles prepared in an entangled pure state like the singlet state. One of them 
remains in laboratory A with Alice and the other is sent to laboratory B to Bob, which 
is far apart creating a noiseless quantum channel. Alice wants to communicate to Bob 
the unknown state of a third particle. She performs a Bell measurement over the two 
particles in her lab and then call Bob to tell him which measurement she did. Thus, 
using this information Bob determines which unitary rotation he must perform to obtain 
the original unknown state of the third particle. After the appearance of this protocol 
many others were proposed and experimentally tested, even considering noisy quantum 
channels~\cite{Bowen:01} as for example thermalized two-qubit Heisenberg 
models~\cite{Yeo:04,Zhou:08,Zhang:07,Kheirandish:08}. 

Thus, considering the standard teleportation protocol using a noisy quantum channel as 
proposed by Bowen and Bose~\cite{Bowen:01}, let's suppose that the initial unknown 
state to be teleported is
\begin{equation}
|\psi(\theta, \phi)\rangle = \cos({\theta}/{2})|0\rangle + 
{e^{i\phi}}\sin({\theta}/{2})|1\rangle
\end{equation}
whose corresponding density operator is
\begin{equation}
\rho_{in} = \frac{1}{2}\left(\mathbb{I} + \hat{n}\cdot\vec{\sigma}\right) ~,
\end{equation}
where $\mathbb{I}$ is the identity matrix, $\hat{n} = (\sin\theta\cos\phi,\sin\theta\sin\phi,\cos\theta)$ and 
$\vec{\sigma} = ({\sigma_{x}},{\sigma_{y}}, {\sigma_{z}})$ are the Pauli matrices. Such density operator 
represents the general state of a single qubit in the Bloch sphere. 

The output state of this protocol is given by~\cite{Bowen:01}
\begin{equation}
{\rho_{out}} = {\sum_{\mu=0}^3}\text{Tr}({\mathcal{K}_\mu}\rho){\sigma_{\mu}}
{\rho_{in}}{\sigma_{\mu}} ~,
\end{equation}
where ${\mathcal{K}_{\mu}} = {\sigma_{\mu}}{\mathcal{K}_0}{\sigma_{\mu}}$, and $\sigma_{0} = \mathbb{I}$, 
$\sigma_{1} = \sigma_{x}$, $\sigma_{2} = \sigma_{y}$ and $\sigma_{3} = \sigma_{z}$, and we choose, without loss of generality, ${\mathcal{K}_0} = |\Psi^{+}\rangle \langle\Psi^{+}|$ based on the possible ground states of the system in Eqs.~\eqref{bellstates}. For 
the present case, we consider the density matrix $\rho$ given in Eq.~\eqref{eq:rhobelldiag} 
representing the noisy teleportation channel where the source of noise is the temperature.

After some simple algebra it is possible to conclude that
\begin{equation}
 \label{eq:productpaulitest}
{\sigma_{\mu}}{\rho_{in}}{\sigma_{\mu}} = \frac{1}{2}\mathbb{I} + 
{\sum_{j=1}^3}\left({\delta_{j\mu}} - \frac{1}{2}\right){n_{j}}{\sigma_{j}} ~.
\end{equation}
Thus, the final density matrix of the output state has the form
\begin{align}
{\rho_{out}} &=  \frac{1}{2}\mathbb{I} + \frac{1}{2}\left(2{p_{\Psi^{+}}} + 2{p_{\Phi^{+}}} - 1 
\right)\sin\theta\cos\phi{\sigma_{1}} \nonumber \\ 
&+ \frac{1}{2} \left(2{p_{\Psi^{+}}} + 2{p_{\Phi^{-}}} - 1\right)\sin\theta\sin\phi{\sigma_{2}} \nonumber \\ 
&+ \frac{1}{2}\left(2{p_{\Psi^{+}}} + 2{p_{\Psi^{-}}} - 1\right)\cos\theta{\sigma_{3}} ~.
\end{align}
where  ${p_{\Phi^{\pm}}} = {\mathcal{Z}^{-1}} {e^{-\beta{E_{\Phi^{\pm}}}}}$ 
and ${p_{\Psi^{\pm}}} = {\mathcal{Z}^{-1}} {e^{-\beta{E_{\Psi^{\pm}}}}} $ are the Boltzmann weights, i.e., the occupation probability of 
each state. With the initial and final states, we may calculate the teleportation fidelity~\cite{Popescu:94,Gisin:96,
Ryszard:96} $f_{\theta, \phi}(\rho_{in}, \rho_{out}) = \langle \psi_{in} | \rho_{out} |\psi_{in}\rangle$. Since the parameters $\theta$ and $\phi$ are unknown, we calculate the average of $f_{\theta, \phi}(\rho_{in}, \rho_{out})$ over the Bloch sphere. Thus,
\begin{equation}
 \label{fidelity}
F = \frac{1}{4\pi} {\int_{0}^{2\pi}}d\phi{\int_{0}^{\pi}}d\theta\sin\theta
{f_{\theta,\phi}}({\rho_{in}},{\rho_{out}}).
\end{equation}
Evaluating $f_{\theta, \phi} (\rho_{in}, \rho_{out})$ and replacing in Eq.~\eqref{fidelity}, 
we then obtain
\begin{equation}
F_{|\Psi^{+}\rangle} = \frac{1}{3} \left(1 + 2{p_{\Psi^{+}}}\right) ~,
\end{equation}
where the index $|\Psi^{+}\rangle$ is just to refer our initial choice of $\mathcal{K}_{0}$. Similarly, 
we can do the same procedure for the others possibles choices of $\mathcal{K}_{0}$ considering our system 
and obtain
\begin{equation}
\label{F_alpha}
F_{|\alpha\rangle} = \frac{1}{3}\left(1 + 2 {p_{\alpha}} \right) ~,
\end{equation}
where $\alpha \in \{ {\Psi^+}, {\Psi^-}, {\Phi^+}, {\Phi^-} \}$. Note that $1/3 < F_{|\alpha\rangle} < 1$. The set of equations ~\eqref{F_alpha}
shows that the teleportation fidelity is directly related to occupation probability of the ground 
state considered.

We have seen early that for different ranges 
of $\Delta$ and $\epsilon$ the ground state of the system changes. Thus, considering this fact, 
the expression for teleportation fidelity will be ${F_{\Phi^{+}}}$, 
${F_{\Phi^{-}}}$, and ${F_{\Psi^{+}}}$ depending on the highest value of fidelity, i.e., the 
value of fidelity related to the highest occupation probability of each eigenstate. In Fig.~\ref{Fidelity} we have 
the function
\begin{equation}
\mathcal{F} = \mbox{max} \left({F_{|\Phi^{+}\rangle}}, {F_{|\Phi^{-}\rangle}}, {F_{|\Psi^{+}\rangle}} \right) ~.
\label{fidelityfigure}
\end{equation}
in terms of $\Delta/{k_B}T$ and $\epsilon/{k_B}T$. The teleportation fidelity is maximum in the same region that 
indicates maximum entanglement ($\mathcal{N} =1$), so that the teleporting process successfully 
occurs. However, it also necessary to know which is the minimal value of fidelity $\mathcal{F}$ to successfully 
occur quantum teleportation. In Refs.~\cite{Pawe:99, Massar:95} the authors show that the minimal 
fidelity $\mathcal{F}$ is inversely proportional to dimension $d$ of Hilbert subspace of a bipartite 
system ($\mathcal{H} = \mathcal{H}_{1} \otimes \mathcal{H}_{2} = \mathcal{C}^{d} \otimes 
\mathcal{C}^{d}$), i.e.,
\begin{equation}
\mathcal{F}_{min} = \frac{2}{1+d} ~.
\end{equation}
For our system $d=2$ and we have that $\mathcal{F}_{min} = 2/3$; in accordance with that result from Ref.~\cite{Verstraete:03}. This boundary is the red line in Fig.~\ref{Fidelity}. It means 
that below this value, the teleporting process ceases. As expected $\mathcal{F}$ 
has the same behavior of figures of negativity and Bell's inequality which means that the non 
possibility to have entangled ground state is, indeed, important. This indicates 
that entanglement and not only non-locality of the ground state is an important condition to 
perform quantum teleportation.
The Fig. \ref{superposition} shows simultaneously the three lines (red lines in Fig. ~\ref{figall}) previously discussed. It is evident in this figure that exists a region (between the green and the red line in Fig. \ref{superposition}) where the states are local but entangled and still the teleportation process successfully occurs.

%%%%%%%%%%%%%%%%%%%%%%%%%%%%%
\section{Discussions}
\begin{figure}[!ht]
\centering
\includegraphics[scale=0.22795]{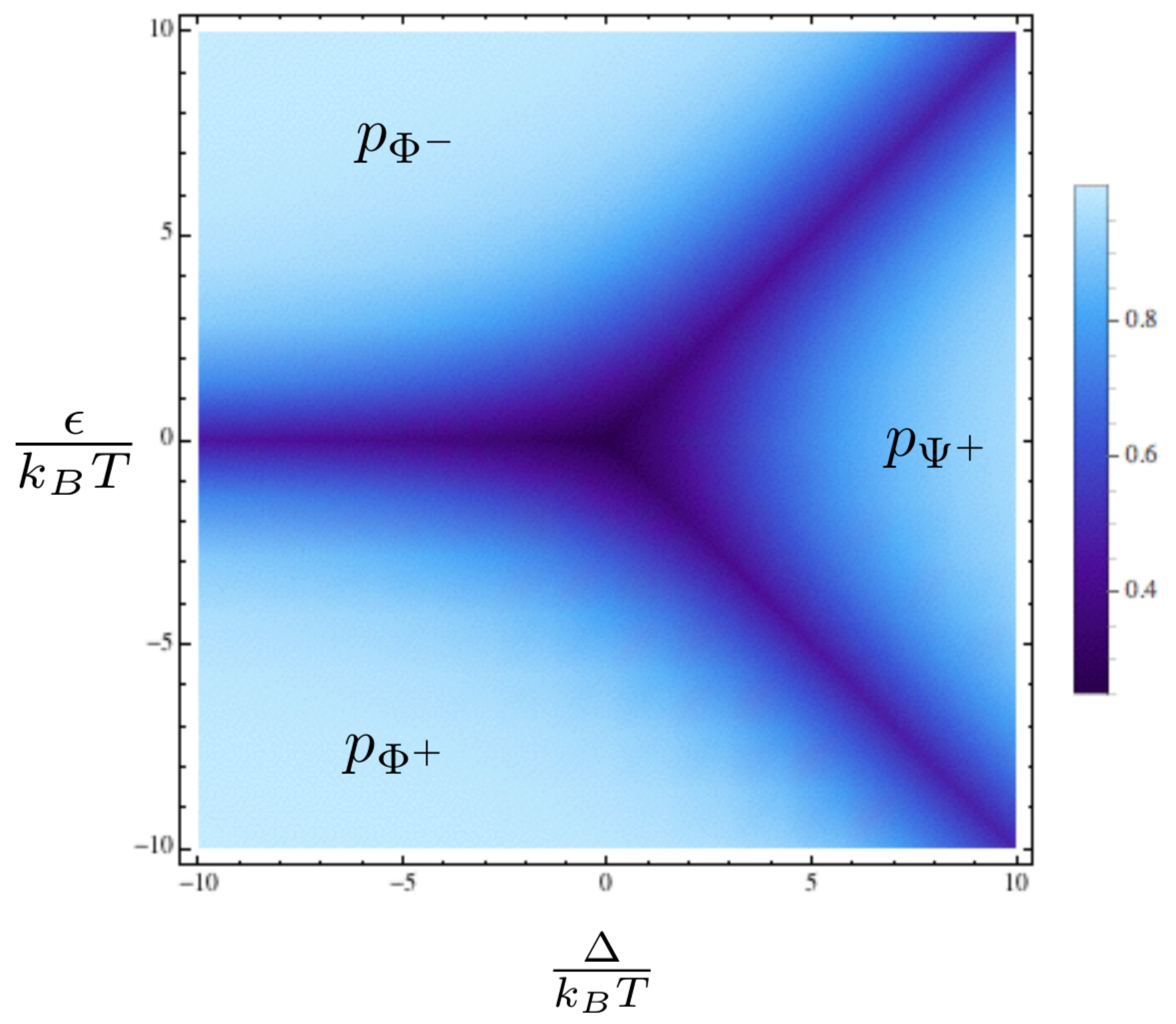}
\caption{(Color on line) Region for function $P$, from Eq.~\eqref{P}, in terms of $\Delta/k_{B}T$ 
and $\epsilon/{k_B}T$. Each region is identified with the dominant Boltzmann's weight.}
\label{pesos}
\end{figure}

%%%%%%%%%%%%%%%%%%%%%%%%%%%%
In Sec.~\ref{entangnonlocal} we study the influence of the dipolar interaction on the thermal entanglement present in a system of two spins-1/2. Thus, our aim is to know how the magnetic anisotropies, represented by the coupling parameters $\Delta$ and $\epsilon$ affect the entanglement, the non-locality and, consequently the teleportation process. In order to observe  the nonlocal correlations we use a measurement based on Bell's inequality called Bell measurement. This quantity is directly related to nonlocal entangled states. In addition, the thermal entanglement of the  system's state was quantified using the negativity. Both quantities were presented in Fig.~\ref{figall}
as a function of $\Delta/{k_B}T$ and $\epsilon/{k_B}T$. In Fig.~\ref{Bell} we have regions were $\mathcal{B}(\rho) > 2$ which means a violation of Bell's inequality and, consequently, the presence of non local states. Maximum entanglement occurs for values in which $\epsilon \to 0$ and $\Delta \gg {k_B}T$ or $\Delta < 0$ and $|\epsilon| \gg {k_B}T$ which corresponds to lightest blue region in Figs. ~\ref{Bell}  and \ref{Negativity}. However, Fig. \ref{Negativity} shows a region, the darkest blue, where negativity is zero and the state of the system is separable. The presence or absence of entanglement is related to the ground state of the system which can change as function of $\epsilon$ and $\Delta$. The change of the ground state can be understood through the variation of the Boltzmann's weights which provide the occupation of the energy levels. Thus, we define the function
\begin{equation}
 \label{P}
P = \max\{{p_{\Phi^{+}}},{p_{\Phi^{-}}},{p_{\Psi^{+}}},{p_{\Psi^{-}}}\} ~,
\end{equation}
i.e., we take the highest value among Boltzmann's weights. Fig.~\ref{pesos} presents Eq.~\eqref{P} in terms of $\Delta/k_{B}T$ and $\epsilon/k_{B}T$ and has the 
same behavior of negativity $\mathcal{N}$ and Bell's inequality violation test $\mathcal{B}$.  In addition, in this figure we identify each region with the dominant Boltzmann's weight. Note that the weight ${p_{\Psi^-}}$ does not appear in the figure due to its lower value compared to the others. The 
coupling parameters $\Delta$ and $\epsilon$ determine which of the four Bell eigenstates is the ground state of the system. It is easy to see by using the set of Eqs.~\eqref{eq:energies0001} and the Boltzmann weights ${p_{\alpha}}$ that in the limit $(\Delta,\epsilon) \ll {k_B}T$ and $\Delta<0$ the ground state is degenerated, this produce a separable mixture of bell states. For the limits 
$\Delta \gg {k_B}T$ the occupation probability is $p_{\Psi^{+}} =1$ while for $\epsilon \gg {k_B}T$, 
$p_{\Phi^{-}} =1$ and for  $- \epsilon \gg {k_B}T$, $p_{\Phi^{+}} =1$. In the last three situations we have a pure and maximally entangled ground state.
From the above results we study the possibility of using such system as a quantum channel in 
order to perform quantum teleportation. The fidelity of teleportation as described in Sec.~\ref{fidelitysection} allow us to know if our system constitute a good channel. As we have seen in that section, depending on the choice of $\mathcal{K}_{0}$ we have different expressions for the fidelity, each one is related to its respective Boltzmann's weight. 

%%%%%%%%%%%%%%%%
\section{Conclusions}
%%%%%%%%%%%%%%%%

To conclude, we demonstrate the effects of the coupling parameters of dipolar interaction between 
two spins $1/2$ particles on the entanglement and non locality of thermal states. This effect is also reflected in the quantum teleportation process. We began by describing the model and showed that the eigenstates of the system may be written in terms of Bell states. We then discussed negativity and the violation of Bell's inequality for this system in order to check the amount of entanglement and the nonlocal correlations present. We showed that the system presents maximally entangled states in the region where  $\epsilon \to 0$ and $\Delta \gg {k_B}T$ or $\Delta < 0$ and $|\epsilon| \gg {k_B}T$
$(\Delta,\epsilon) \gg {k_B}T$, according to negativity. 

We then proceeded by providing an application for such model in a quantum teleportation process. 
We evaluated the fidelity of teleportation $\mathcal{F}$, which depends on the probability of  occupation of each Bell state, and showed that this quantity is maximum when entanglement is maximum and the teleportation process ceases when there is no entanglement as expected. For this system, the minimal value of $\mathcal{F}$ to successfully occurs quantum teleportation is $2/3$. As a remarkable result we found that, even without violation of the Bell inequality, the 
teleportation successfully occurs as shows the Fig.~\ref{superposition}. Thus, for the present system, all the entangled states are useful for teleportation.

Finally, this study provides relevant insights into how the parameters of dipolar interaction coupling 
between two spins $1/2$ particles affect entanglement, violation of Bell's 
inequality, and teleportation fidelity. Thus, we advocate that this model of dipolar interaction can be used as a good quantum channel to promote quantum communication.

%%%%%%%%%%%%%%%%%%%%%%%%%%%%%%%%%%%%%%%%%%%%%%%%%%%%%%%%
\section*{ACKNOWLEDGEMENTS}
C. S. Castro thanks to L. Justino and A. M. Souza for discussions on Bell's inequalities. The 
authors gratefully acknowledgement financial support from CNPq, CAPES, FAPERJ, grant 2014/00485-7 S\~{a}o Paulo Research Foundation (FAPESP), and 
INCT-Informa\c{c}\~{a}o Qu\^{a}ntica. 
%%%%%%%%%%%%%%%%%%%%%%%%%%%%%%%%%%%%%%%%%%%%%%%%%%%%%%%

\bibliographystyle{apsrev4-1}
%\bibliography{bib}


\end{document}